\documentclass[letterpaper]{article} 
\usepackage{aaai23}  
\usepackage{times}  
\usepackage{helvet}  
\usepackage{courier}  
\usepackage[hyphens]{url}  
\usepackage{graphicx} 
\usepackage{natbib}  
\usepackage{caption} 
\usepackage{algorithm}
\usepackage{algorithmic}

\usepackage{newfloat}
\usepackage{listings}
\DeclareCaptionStyle{ruled}{labelfont=normalfont,labelsep=colon,strut=off} 
\floatstyle{ruled}
\newfloat{listing}{tb}{lst}{}
\floatname{listing}{Listing}
\nocopyright  

\usepackage{amsmath}
\usepackage{amsfonts}
\usepackage{amsthm}
\usepackage{subcaption}
\usepackage{xcolor}

\newcommand\extended[1]{\textcolor{black}{#1}}

\title{Estimating Geographic Spillover Effects of COVID-19 Policies\\From Large-Scale Mobility Networks\thanks{\extended{This is the extended version of a paper accepted to AAAI'23.}}}
\author{
    Serina Chang\textsuperscript{\rm 1}, Damir Vrabac\textsuperscript{\rm 1}, Jure Leskovec\textsuperscript{\rm 1}, Johan Ugander\textsuperscript{\rm 2}
}
\affiliations{
    \textsuperscript{\rm 1}Department of Computer Science, Stanford University\\
    \textsuperscript{\rm 2}Department of Management Science \& Engineering, Stanford University\\
    \{serinac, dvrabac, jure, jugander\}@stanford.edu
}

\begin{document}

\maketitle

\begin{abstract}
Many policies in the US are determined locally, e.g., at the county-level. Local policy regimes provide flexibility between regions, but may become less effective in the presence of geographic spillovers, where populations circumvent local restrictions by traveling to less restricted regions nearby.
Due to the endogenous nature of policymaking, there have been few opportunities to reliably estimate causal spillover effects or evaluate their impact on local policies.
In this work, we identify a novel setting and develop a suitable methodology that allow us to make unconfounded estimates of spillover effects of local policies. Focusing on California’s Blueprint for a Safer Economy, we leverage how county-level mobility restrictions were deterministically set by public COVID-19 severity statistics, enabling a regression discontinuity design framework to estimate spillovers between counties. We estimate these effects using a mobility network with billions of timestamped edges and find significant spillover movement, with larger effects in retail, eating places, and gyms.
Contrasting local and global policy regimes, our spillover estimates suggest that county-level restrictions are only 54\% as effective as statewide restrictions at reducing mobility. 
However, an intermediate strategy of macro-county restrictions---where we optimize county partitions by solving a minimum $k$-cut problem on a graph weighted by our spillover estimates---can recover over 90\% of statewide mobility reductions, while maintaining substantial flexibility between counties.
\end{abstract}

\section{Introduction}
Many policies in the United States---COVID-19 restrictions, environmental regulations, and laws controlling the sales of e-cigarettes, firearms, and controlled substances---are determined at the state- or county-level.
Local policy regimes provide flexibility between regions, allowing policymakers to set regulations depending on local circumstances (e.g., COVID-19 severity) and the preferences of their constituents (e.g., on gun control).
On the other hand, allowing policies to be set locally often results in differing levels of restrictiveness between neighboring regions. These differences can lead to \textit{geographic spillovers}, where populations circumvent restrictions by traveling to less restricted regions nearby. 
Spillovers risk undermining the efficacy of local policies; for example, if banned goods are imported across state borders or if, during the pandemic, individuals in counties under lockdown continue to visit places in neighboring counties.
Furthermore, spillovers can affect important downstream consequences. For example, the movement of individuals from more restricted (and possibly more infected) regions to less restricted (and possibly less infected) regions during the pandemic could result in greater overall spread of the virus.

However, there are few opportunities to reliably estimate causal spillover effects.
Researchers cannot run experiments to randomly assign policies to states and counties, and causal identification is difficult in most observational studies, due to the presence of confounders. For example, attempting to study the effects of COVID-19 restrictions (e.g., closing restaurants) on mobility patterns introduces potential confounding covariates that predict both the treatment and the outcome, such as current COVID-19 severity in the region and the population's demographics. 
Prior work has attempted to address these confounders by controlling for them, but there could always be unobserved or unknown confounders that bias causal estimates.
Furthermore, the decentralized nature of policymaking that gives rise to potential spillovers also often results in varying policy definitions and implementations across regions.
This heterogeneity makes it difficult to define a consistent treatment whose effects we can measure.

In this work, we introduce a setting in which we can make unconfounded estimates of the spillover effects of consistent policies.
We focus on California's Blueprint for a Safer Economy, a statewide policy framework that determined weekly county-level mobility restrictions for all 58 counties in California from August 2020 to June 2021.
The Blueprint consisted of four tiers that corresponded to policies of decreasing restrictiveness. 
At the start of each week, each county’s tier was determined based on that county's COVID-19 metrics (case rate and test positivity) in the preceding weeks.
The California Blueprint presents a unique opportunity for studying spillover for three reasons: (1) neighboring counties were frequently in differing tiers, enabling the analysis of spillovers from more restricted to less restricted counties; (2) tiers were defined in the same way across counties, yielding a consistent treatment; (3) tiers were deterministically assigned at the thresholds of COVID-19 metrics. These three ingredients allow us to develop a causal inference framework based on regression discontinuity design to make unconfounded estimates of spillover effects.

To capture spillover, we focus on cross-county mobility in a large-scale mobility network. 
Our network is a dynamic bipartite graph that represents the weekly movements of individuals from census block groups (CBGs) to specific points-of-interest (POIs) such as restaurants and grocery stores. 
Our objective is to estimate the effect of pairwise county tiers on the number of visits from each CBG to POI. 
The mobility network for California contains around 23,000 CBGs and 130,000 POIs, with nearly 3 billion edges per week. 
We use stochastic gradient descent, with loss-corrected negative sampling, to make estimation computationally feasible in this large-scale setting.
Studying mobility patterns at the POI-level enables us to estimate heterogeneous treatment effects for POI categories; this ability is particularly relevant since tier restrictions were often industry-specific.

Finally, our spillover estimates allow us to quantify the cost of spillovers on policies across spatial scales. 
In the presence of spillovers, we find that county-level restrictions are, on average, only 54\% as effective as statewide restrictions at reducing mobility.
However, intermediate strategies of macro-county restrictions---when counties are grouped intelligently---can balance the trade-off between the policy flexibility and efficacy.
We show that finding the most effective county partition for a given spatial granularity is equivalent to solving a minimum $k$-cut problem on an undirected county graph weighted by our spillover estimates.
Using this approach, we identify macro-county restrictions that recover over 90\% of statewide mobility reductions, while maintaining substantial flexibility between counties.

In summary, our contributions are as follows:
\begin{itemize}
    \item \textbf{Setting:} we identify a novel setting for studying spillovers where the same set of policies was applied with the same thresholds to many areas;
    \item \textbf{Methods:} we develop a regression discontinuity (RD) design framework that allows us to make unconfounded estimates of heterogeneous spillover effects in this setting, estimated over a large-scale mobility network containing billions of edges;
    \item \textbf{Analyses:} we demonstrate significant spillover effects in many POI groups and evaluate the costs of these spillovers on policies across spatial scales.
\end{itemize}
In a complex, interconnected world with few opportunities to reliably estimate policy effects, our work is among the first to identify a setting where spillovers can be rigorously estimated and to develop an appropriate methodology to estimate and evaluate the effects of spillovers.\footnote{The code to run our experiments and regenerate figures is available at \url{https://github.com/snap-stanford/covid-spillovers}. 
We also provide our constructed $Z$ variables (Section \ref{sec:causal}) that can be used with RD design to estimate the effects of the California Blueprint tiers on spillovers and other outcomes.
}


\section{Related Work}
Spillovers often arise from decentralized policymaking for interconnected regions. 
For example, \citet{sigman2005environmental} finds that water quality is lower at stations downstream of states that are authorized to control their own water programs, since they ``free-ride.'' \citet{coates2017guns} show that in states with stronger gun laws, there is an increased likelihood of gun imports from states with weaker gun laws. 
\citet{bronars1998guns} show that while a concealed handgun law led to a reduction in crime in the state, it also led to an increase in crimes in neighboring states, suggesting that criminals were crossing borders.
\citet{hao2017marijuana} find that legalization of recreational marijuana in a state leads to an increase in marijuana-related arrests in bordering states. 
Spillovers also arise in online contexts, where instead of crossing geographic borders, users can migrate across platforms if they are banned on one platform; furthermore, levels of toxicity and radicalization are sometimes higher on the new, often less regulated platforms, compromising the efficacy of the original content moderation \citep{ribeiro2021content, ali2021deplatforming}. 

In the context of COVID-19, prior research has mostly focused on the direct effects of policies on population health or behavior, without explicitly modeling spillovers \citep{chernozhukov2021causal, nguyen2020mobility, brauner2020science}. \citet{chandrasekhar2021interacting} investigate disease spillovers between interconnected regions in a model-based setting and \citet{holtz2020interdependence} provide early evidence of mobility spillovers, showing that a state's population reduced its own mobility when neighboring states implemented shelter-in-place policies. Most related to our work is \citet{zhao2021spillovers}, who use a difference-in-difference approach to estimate the effects of COVID-19 policies on mobility and provide evidence of spillovers in cross-state travel. We build on this work by addressing two primary limitations of their study: first, the authors note that their estimates could be confounded by unobserved, time-varying factors; other research on spillovers also suffers from potential confounding, using difference-in-difference approaches \citep{hao2017marijuana, holtz2020interdependence} or regressions \citep{coates2017guns, sigman2005environmental, bronars1998guns}. 
Second, in order ``to create sufficient statistical power to identify causal effects,'' the authors collapse different policy interventions into general policy ``types'' (e.g., resuming dine-in and lifting gathering restrictions are both counted as reopening), which violates assumptions of consistent treatment.

In contrast with prior work, we are able to identify unconfounded spillover estimates for a single set of policies by applying our RD-based framework to California's Blueprint for a Safer Economy. 
Furthermore, by estimating effects on the CBG-POI network, our model enables the analysis of counterfactual fine-grained mobility patterns under different pandemic policies. 
Understanding mobility patterns has been essential to controlling the spread of COVID-19 \citep{buckee2020aggregated}, and many researchers rely on fine-grained mobility data to model the effect of mobility on the spread of the virus \citep{badr2020mobility, chinazzi2020, kraemer2020mobility, chang2020covid, chang2021kdd, nouvellet2021reduction}.
Our model furthers such analyses by investigating the complex effects of policy interventions on mobility, closing the gap from policy to behavior to COVID-19 outcomes.
\section{Data} \label{sec:data}
\begin{figure}
    \centering
    \includegraphics[width=\linewidth]{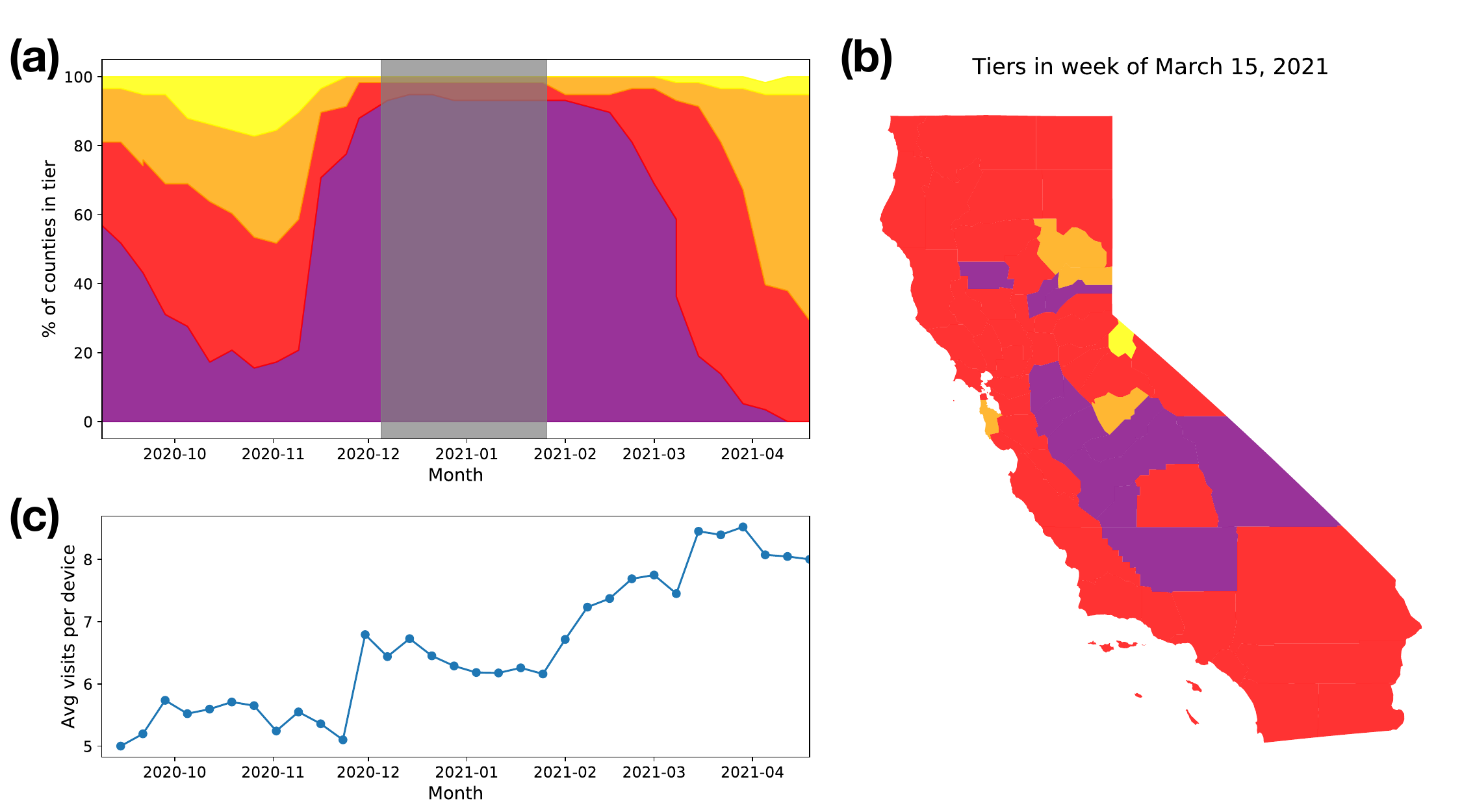}
    \caption{Primary data sources. (a) Percentage of California counties in Blueprint tiers---purple, red, orange, and yellow---over time (greyed-out period represents Regional Stay-At-Home Order); (b) Tiers in the week of March 15, 2021; (c) Average weekly visits per device over time.}
    \label{fig:data_plots}
\end{figure}

\paragraph{California Blueprint for a Safer Economy.}
The Blueprint was implemented for all 58 counties in California from August 30, 2020 to June 15, 2021. It consisted of four tiers: purple (``widespread''), red (``substantial''), orange (``moderate''), and yellow (``minimal''). These tiers corresponded to mobility policies of decreasing restrictiveness; for example, in the purple tier, most non-essential indoor businesses were closed, while in yellow, they could be open with modifications.
We use the archived data sheets from the California Department of Public Health (CDPH), which provide detailed documentation of every county's weekly tier assignment and the COVID-19 metrics used to make those assignments.\footnote{\url{https://www.cdph.ca.gov/Programs/CID/DCDC/Pages/COVID-19/CaliforniaBlueprintDataCharts.aspx}}
In Figure \ref{fig:data_plots}a, we visualize the progression of counties through tiers over time; we grey out the period from December 5, 2020 to January 25, 2021, during which most of the state was under a Regional Stay-At-Home Order \citep{cdph2020stayathome}. 
We can see that counties generally moved through similar tiers at similar times, which is expected, since COVID-19 severity was correlated across counties. 
However, in many weeks, we also see substantial representation from at least two different tiers. For example, in the week of March 15, 2021, there were 11 counties in the purple tier, 42 in the red tier, 4 in the orange tier, and 1 in the yellow tier (Figure \ref{fig:data_plots}b). Many of these differing tiers appeared between adjacent counties, enabling the analysis of spillovers across county borders.

\paragraph{Mobility network.} 
We use data from SafeGraph, a company that anonymizes and aggregates location data from mobile apps.
For each POI, SafeGraph provides weekly estimates of where visitors are coming from, aggregated over CBGs.\footnote{\url{https://docs.safegraph.com/docs/weekly-patterns}} 
This creates a dynamic, bipartite graph between CBGs and POIs, where an edge weight $Y_{ijw}$ represents the number of visits recorded by SafeGraph from CBG $c_i$ to POI $p_j$ in week $w$. SafeGraph also reports how many devices they recorded in each CBG and week. Incorporating device counts into our model allows us to account for varying coverage across CBGs and over time.

In Figure \ref{fig:data_plots}c, we show the average number of weekly visits recorded per device over time, aggregated over the entire CBG-POI network for California. We see that visits increased post-Regional Stay-at-Home as Blueprint tiers decreased in restrictiveness. 
However, various latent variables could explain this correlation, such as reduced COVID-19 severity leading to less restrictive tiers
and less fear of visiting places. 
Thus, it is necessary to develop a robust causal framework that allows us to disentangle tier effects from confounders, which we describe in the following section.


\section{Causal framework} \label{sec:causal}
\begin{figure*}
    \centering
    \includegraphics[width=\linewidth]{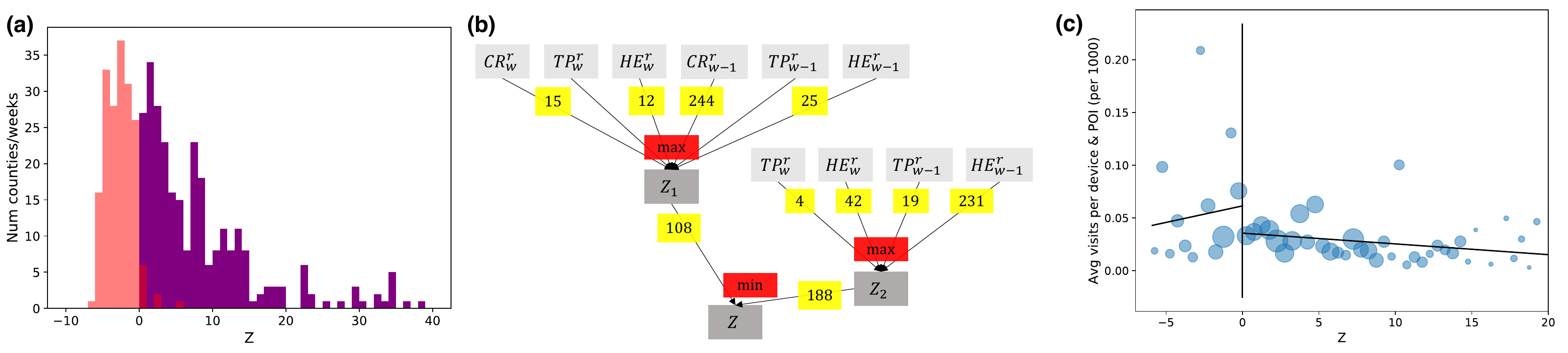}
    \caption{Visualizing our $Z$ variable. (a) $Z$ almost perfectly separates counties in the purple and red tiers. 
    (b) Triggering patterns for $Z$ (among large counties compliers).
    (c) Cross-county mobility vs $Z$. 
    All source counties are in purple ($0 \leq Z \leq 5$) and the x-axis represents the target county's $Z$. Black lines represent linear fits and dots are average outcomes per bin (size represents the number of data points in the bin). We observe a discontuinity at $Z=0$, when the target county changes from purple to red.}
    \label{fig:z_viz}
\end{figure*}

To capture spillovers, our objective is to estimate the effect of \textit{pairwise} tiers on cross-county mobility. 
The key to our causal framework is that we can utilize RD design, which is widely recognized as ``one of the most credible non-experimental strategies for the analysis of causal effects'' \citep{cattaneo2020intro}. In a typical RD design, units are assigned to the treatment or control condition according to an exogenously determined threshold of a single continuous variable, known as the assignment variable (or running or forcing variable). 
Researchers can then compare the outcomes for units just below the threshold to units just above the threshold to estimate the \textit{local} causal effect of treatment.
A primary advantage of RD design is that it achieves unconfoundedness, without needing to control for all possible confounders. 
This is because the unconfoundedness assumption is met: treatment assignment is conditionally independent of potential outcomes, given covariates \citep{imbens2008rd}. 
This assumption is clearly met in RD design, since treatment assignment is determined by the assignment variable, and so, conditioned on covariates, there is no variation in treatment. 

Our problem generally fits RD design, since Blueprint tiers were assigned at the thresholds of continuous COVID-19 metrics. We focus on the threshold between the purple and red tiers, since they were the adjacent pair with the most support. 
However, we need to extend generic RD design in two ways: (1) to account for multiple assignment variables, since tiers were assigned based on numerous COVID-19 metrics, (2) to account for multiple treatment conditions, since we are considering pairwise tiers as our treatment. We describe our approach in the following sections.

\paragraph{Assigning Blueprint tiers.}
First, let us focus on the problem of determining a single county's tier, $T_{iw}$, from its COVID-19 metrics.
Tier assignments depended on three metrics: adjusted case rate, test positivity rate, and a health equity metric, which was the test positivity rate in the most disadvantaged quartile of neighborhoods \citep{cdph2021healtheq}.
To advance to a less restricted tier, counties needed to meet the criteria for movement for two consecutive weeks \citep{cdph2021blueprint}. For a large county (population over 106,000), the criteria to move from purple to red could be met in two ways: (1) by meeting the thresholds for the red tier for all three metrics, (2) by meeting the thresholds for test positivity and health equity for the \textit{orange} tier, thus exchanging adjusted case rate for more stringent thresholds on the other two.
\extended{Small counties (population under 106,000) only had one possible path, which was to meet the adjusted case rate and test positivity thresholds for the red tier.}
Small counties were not required to meet the health equity thresholds, but needed to demonstrate their commitment to equity through other plans \citep{cdph2021healtheq}.
For most of the duration of the Blueprint, the purple-red threshold for adjusted case rate was 7 per 100,000 and 8\% for test positivity and health equity (and 5\% for the red-orange threshold). The purple-red threshold for adjusted case rate was increased to 10 per 100,000 on March 12, 2021, after 2 million vaccines had been administered statewide \citep{calmatters2021vaccinegoal}.

\paragraph{Constructing a single assignment variable $Z$.}
We take a \textit{centering} approach to RD design with multiple assignment variables \citep{wong2013centering}. 
That is, we first center each of the assignment variables by subtracting their respective thresholds, then apply min/max aggregations to the centered variables in order to construct a new assignment variable $Z$ that can singly determine a unit's treatment.
More formally, we design a function $f: \mathbb{R}^m \rightarrow \mathbb{R}$ that maps a county’s $m$ COVID-19 metrics to a single continuous variable, $Z_{iw}$. For a large county, the $m$ metrics include the county's adjusted case rate ($CR$), test positivity ($TP$), and health equity metric ($HE$) from the preceding two weeks; for a small county, only adjusted case rate and test positivity. Our mapping $f$ satisfies the key property that $Z_{iw} < 0$ if and only if the county was assigned to the red tier.

Let $CR_{iw}^{\textrm{red}}$ represent the adjusted case rate for the county in week $w$ with the purple-red threshold subtracted, and let us define other terms similarly. We construct $Z_{iw}$ for large counties as follows:
\begin{align}
    Z_{1iw} &= \max(CR_{iw}^{\textrm{red}}, TP_{iw}^{\textrm{red}}, HE_{iw}^{\textrm{red}}, CR_{iw-1}^{\textrm{red}}, TP_{iw-1}^{\textrm{red}}, HE_{iw-1}^{\textrm{red}})\\
    Z_{2iw} &= \max(TP_{iw}^{\textrm{orange}}, HE_{iw}^{\textrm{orange}}, TP_{iw-1}^{\textrm{orange}}, HE_{iw-1}^{\textrm{orange}})\\
    Z_{iw} &= \min(Z_{1iw}, Z_{2iw}).
\end{align}
For small counties, we only have
\begin{align}
    Z_{iw} &= \max(CR_{iw}^{\textrm{red}}, TP_{iw}^{\textrm{red}}, CR_{iw-1}^{\textrm{red}}, TP_{iw-1}^{\textrm{red}}).
\end{align}
In Figure \ref{fig:z_viz}a, we show that our new $Z$ variable almost perfectly separates the counties in the purple and red tiers. 
Over the 9-week period from February 1 to March 29, 2021, there were 480 counties/weeks in the purple or red tier, and 471 of them follow that $Z_{iw} < 0$ if and only if the county is in the red tier. 
We manually check the non-compliers and find that they were cases of counties, mostly small, that were allowed to remain in the red tier upon special request, as noted in the CDPH documentation.

To interpret our new $Z$ variable, we also analyze its ``triggering'' patterns; that is, for each min/max aggregation, which input is the minimum or maximum (Figure \ref{fig:z_viz}b). 
\extended{Since $Z < 0$ moves the county into the red (less restricted) tier, a maximum can be interpreted as holding the county back and a minimum as improving the county's prospects.}
For large counties, we find that the most frequent maximum for the first criteria $Z_1$ is the adjusted case rate from week $w-1$. For the second criteria $Z_2$, the most frequent maximum is the health equity metric from week $w-1$. This reflects trends from this time period: COVID-19 severity was improving over time, so week $w-1$ tended to have higher rates than week $w$, and health equity (i.e., test positivity in the most disadvantaged quartile) tended to be worse than the overall test positivity. 
Interestingly, we also find that $Z_2$ triggers more often than $Z_1$, when taking the minimum between them. This indicates that this alternative path---meeting more stringent test positivity and health equity thresholds and dropping adjusted case rate---substantially helped counties move toward less restricted tiers.

\paragraph{RD design with pairwise treatments.}
We can now formulate an RD design problem where treatment (purple/red tier) is assigned at the threshold of a single continuous variable ($Z$). Since we are interested in spillover effects in this work, we use \textit{cross-county} mobility as our outcome. However, our RD framework is general and could be applied to study the effects of Blueprint tiers on a variety of outcomes, such as mask-wearing rates, vaccination rates, and COVID-19 cases and deaths.

With cross-county mobility as our outcome, our treatment becomes pairwise to capture the tier of each county, and we have four treatment conditions: $PP, PR, RP$ and $RR$, where $P$ and $R$ represent the purple and red tiers, respectively.
We are particularly interested in the difference between $PP$ and $PR$, since this difference indicates whether individuals from a restricted county will increase their visits to another county when that other county becomes less restricted.
In Figure \ref{fig:z_viz}c, we illustrate this comparison. We consider all source counties that were in the purple tier and plot their mobility to target counties that were either in the purple or red tier. The x-axis represents $Z$ for the target county, so that the region to the left of $Z=0$ represents the $PR$ condition and the region to the right represents $PP$. We see a discontinuity in visits at $Z=0$, indicating that there is indeed a local effect on cross-county visits when a neighboring county changes from more to less restricted.
In the following section, we estimate this effect more precisely by defining a zero-inflated Poisson regression model that we fit to the rich CBG-POI mobility network with covariates.

\paragraph{Poisson regression model.}
We define a Poisson regression model to describe visits from CBGs to POIs. 
For a given CBG $c_i$, POI $p_j$, and week $w$, the Poisson rate $\lambda_{ijw}$ is
\begin{align}
    \lambda_{ijw} &= \exp(\beta_0 + \beta_1 Z_{iw} + \beta_2 Z_{jw} +  \mathbf{\beta_3}^T\mathbf{X}_{ijw} + \beta_{T_{iw}, T_{jw}}),
\end{align}
where the $\beta$ terms are model parameters, $Z_{iw}$ and $Z_{jw}$ represent the $Z$ variables for $c_i$'s and $p_j$'s counties in this week, $T_{iw}$ and $T_{jw}$ describe their respective tiers, and $\mathbf{X}_{ij}$ contains other covariates. Those covariates include the distance between the POI and CBG, SafeGraph's CBG device count in that week, CBG demographics from US Census, and POI attributes (area in square feet, NAICS code).
Spillover effects are captured in the difference between the $\beta_{T_{iw}, T_{jw}}$ terms: for example, $\exp(\beta_{PR} - \beta_{PP})$ represents the multiplicative increase in visits when a POI changes from the purple to red tier, while the CBG remains in purple.
\extended{To capture heterogeneous treatment effects, we learn separate $\beta_{T_{iw}, T_{jw}}$'s for different POI groups (Table \ref{tab:poi-groups}). We also learn separate $\beta_1$'s and $\beta_2$'s for our four different constructions of $Z$ that reflect two binary dimensions: 1) large vs. small county, 2) before vs. after March 12, 2021, when the statewide vaccine goal was met and the adjusted case rate threshold was increased.}

The CBG-POI network is very large, with billions of edges, but over 99\% of the edges represent zero visits. Thus, we zero-inflate our Poisson model, based on the notion that observed zeros in zero-heavy data may represent actual preferences, but could also reflect lack of exposure \citep{liu2017zeroinf}, i.e., the CBG had never heard of the POI. We represent each number of visits $Y_{ijw}$ as drawn from a mixture of a Poisson($\lambda_{ijw}$) and $\delta_0$ (a point mass on 0), with mixing parameter $\pi_{ij}$. We assume the likelihood of exposure is inversely proportional to the distance $d_{ij}$ between the CBG and POI and define $\pi_{ij} = \frac{1}{1 + \alpha_1 d_{ij}^{\alpha_2}}$, where the $\alpha$ terms are learned.
Then, our generative model for $Y_{ijw}$ is
\begin{align}
    b_{ijw} &\sim \textrm{Bern}(\pi_{ij}) \\
    Y_{ijw} &\sim \begin{cases}
            \delta_0\textrm{ if $b_{ijw} = 0$,}\\
            \textrm{Poisson}(\lambda_{ijq}),\textrm{ otherwise.}
        \end{cases} 
\end{align}
In this mixture, the likelihood of a single data point given model parameters $\theta$ is
\begin{align}
    \Pr(Y_{ijw} = y | \theta) = \begin{cases}
        (1 - \pi_{ij}) + \pi_{ij} e^{-\lambda_{ijw}},\textrm{ if $y = 0$}\\
        \pi_{ij} \frac{\lambda_{ijw}^y e^{-\lambda_{ijw}}}{y!},\textrm{ otherwise.}\\
    \end{cases}
\end{align}
We fit our model using gradient descent, with negative log likelihood as our model loss.

\paragraph{Data filtering and bandwidth selection.}
We focus our experiments on the 9-week period following the Regional Stay-At-Home Order, during which we could almost perfectly separate the purple and red tiers with our $Z$ variable (Figure \ref{fig:z_viz}a). 
\extended{We keep all CBGs with at least 50 non-zero visits (to any POIs) during this period and POIs with at least 30 non-zero visits (from any CBGs), so that we focus on CBGs and POIs for which SafeGraph has more reliable coverage; this filtering leaves 22,972 CBGs and 128,655 POIs.}
Due to the specifics of our RD-based analysis, we cannot keep every CBG-POI pair from every week. 
First, we do not fit the model on data from the week of March 8, 2021, since the purple-red threshold for adjusted case rate was changed in the middle of the week (due to the statewide vaccine goal being met). 
In the remaining 8 weeks, we keep all data points that meet the following criteria:
\begin{itemize}
    \item CBG $c_i$ and POI $p_j$ lie in adjacent counties,
    \item $T_{iw}$ and $T_{jw}$ are both in the purple or red tier,
    \item Both are compliers, i.e., $T_{iw}$ is red if and only if $Z_{iw} < 0$, and likewise for $T_{jw}$,
    \item $Z_{iw}$ and $Z_{jw}$ both lie within a bandwidth $h$ of 0.
\end{itemize}
\extended{In total over the 8 weeks, we keep 1.4 billion data points after filtering (Table \ref{tab:data_size}).} 

We only keep data points that fall within the bandwidth since our goal is to estimate the \textit{local} effect of changing tier pairs at the purple-red threshold ($Z=0$). 
By requiring both $Z_{iw}$ and $Z_{jw}$ to fall within the bandwidth, we interpret our resulting parameters as estimated effects at the \textit{joint} cutoff, when both the CBG and the POI are at the threshold.\footnote{Alternatively, RD design with multiple assignment variables can estimate effects along the threshold frontiers, i.e., varying one assignment variable while fixing the other one at its threshold \citep{papay2011multiple}. For simplicity, we focus on effects at the joint threshold.}
Bandwidth selection introduces a bias-variance trade-off, with larger bandwidths corresponding to greater bias but reduced variance. We err on the side of larger bandwidths in this work, out of concern for variance. Even though we have over a billion data points, our assignment variable $Z$ only varies at the level of counties and, thus, bandwidths that are too small could lead to very few counties represented, particularly for the $PR$ or $RP$ treatment conditions, which appear less often. We choose $h=5$, which keeps most of the counties in the red tier, but drops many of the counties in purple (Figure \ref{fig:z_viz}a). 
We show in the Appendix that each treatment condition is well-represented at this bandwidth, with a diversity of county pairs \extended{(Table \ref{tab:h5_data_stats})}.
Furthermore, we conduct sensitivity analyses with $h=4$ and $h=6$ and show that results remain highly similar \extended{(Figure \ref{fig:bandwidth_results})}.

\paragraph{Loss-corrected negative sampling.}
To make estimation computationally feasible in this large-scale setting, we perform negative sampling. 
Specifically, for each zero data point $(i, j, w)$, we define its sampling probability $s_{ijw}$ as inversely proportional to the distance between the CBG and POI  ($s_{ijw} \propto \frac{1}{1 + d_{ij}}$). We do this to upweight ``hard'' negative samples; that is, since far-apart CBGs and POIs are highly unlikely to have any visits, the model learns more from nearby CBGs and POIs with zero visits.
However, a unique aspect of our problem---which does not typically appear in other machine learning prediction problems where negative sampling might be used, such as link prediction or learning word embeddings---is that because we seek to interpret the model parameters as effect sizes, our learned model parameters need to be unbiased estimates of the model parameters when learned on the full data.
Left uncorrected, negative sampling biases our model parameters by greatly reducing the number of zeros in the training data.

In the Appendix we show that by weighting each sampled zero data point by $\frac{1}{s_{ijw}}$ when computing the overall loss (negative log likelihood), our stochastic gradient (which is stochastic from sampling) forms an unbiased estimate of the true gradient, which ultimately guarantees unbiased parameter estimates assuming proper model specification. We also show that upweighting harder negative samples, as well as increasing the size of the sample, decreases the variance of the stochastic gradient, providing formal validation of these techniques.
In our experiments, we retain 2\% of the zero data points, with sampling probabilities inversely weighted by distance. We verify that after incorporating our loss corrections, different negative sampling schemes arrive at the same average parameters, but distance-weighting and larger samples decrease variance. The agreement between the estimates from different negative sampling schemes is consistent with the underlying model being properly specified.

\paragraph{Uncertainty quantification with bootstrapping.}
We run 30 trials, where in each trial, we perform negative sampling on the zero data points and we sample $N_{nnz}$ non-zero data points with replacement, where $N_{nnz}$ is our total number of non-zero data points. For a given estimand, such as $\tau_{PR} = \exp(\beta_{PR} - \beta_{PP})$, we compute its 95\% confidence interval as $\bar{\tau}_{PR} \pm 1.96 \cdot \hat{\sigma}_{\tau_{PR}}$, where $\bar{\tau}_{PR}$ and $\hat{\sigma}_{\tau_{PR}}$ are its sample mean and standard deviation over trials, respectively. This procedure captures uncertainty from the data and from negative sampling, although we show that, given our chosen negative sampling scheme, the former accounts for the vast majority of the variance \extended{(Figure \ref{fig:neg_sample_cis})}.
\section{Results}
\label{sec:results}
\paragraph{Spillover estimates.}
\extended{We learn heterogeneous effects for different POI groups, where we consider all ``top''-categories (first 4 digits of the POI's NAICS code) with at least 1000 POIs.\footnote{Following prior work using SafeGraph data \citep{chang2020covid}, we drop the category ``Elementary and Secondary Schools'' due to poor coverage of children from cell phone data.}
We also include separate effects for 4 ``sub''-categories (first 6 digits) of interest, all of which have over 1000 POIs. In Table \ref{tab:poi-groups}, we provide the NAICS codes, descriptions, and number of POIs in each POI group.}

\begin{figure}
    \centering
    \includegraphics[width=\linewidth]{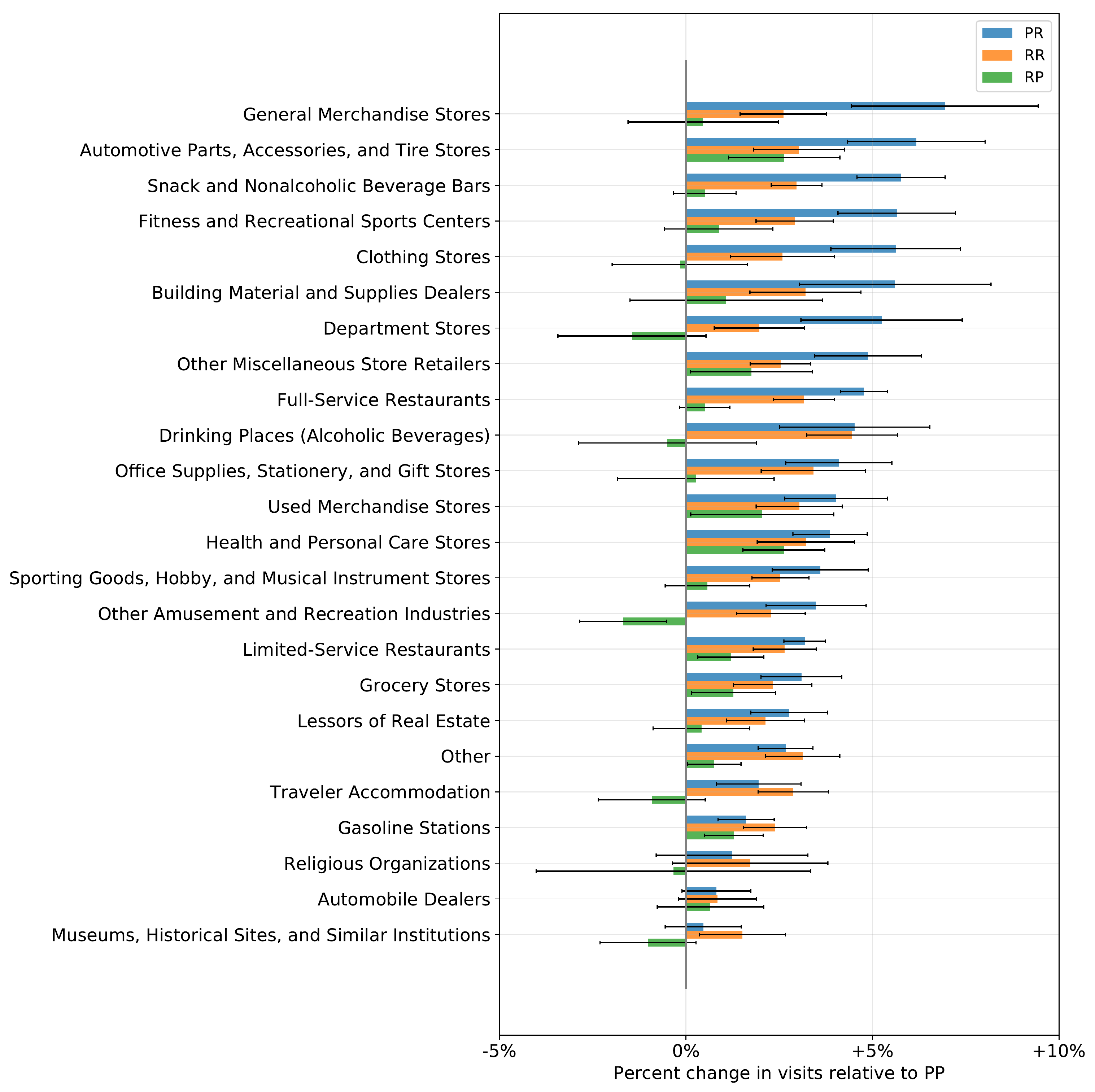}
    \caption{Estimated spillover effects across POI groups, with 95\% confidence intervals.}
    \label{fig:spillover_results}
\end{figure}

We present our spillover results in Figure \ref{fig:spillover_results}.
First, we find significant positive $PR$ effects in 21 out of 24 groups (all results remain significant with Bonferroni correction). That is, visits from the CBG increase significantly when the POI's county changes from purple to red, while the CBG's county remains in purple. This indicates spillovers, as people from more restricted counties spill over in less restricted, adjacent counties. 
Furthermore, we observe varying effect sizes; for example, with larger effects in retail (General Merchandise Stores, Automotive Stores, Clothing Stores, Building Material and Supplies Dealers, Department Stores), most eating places (Snack Bars, Full-Service Restaurants, Drinking Places), and gyms (Fitness and Recreational Sports Centers). 
Smaller effects are in essential retail (Grocery Stores, Gas Stations), hotels (Traveler Accommodation), malls (Lessors of Real Estate), museums, historical sites, and nature parks (Museums, Historical Sites, and Similar Institutions).
\extended{This heterogeneity in effect size may partially reflect differences across sectors in tier restrictions. For example, essential retail, hotels, and malls remained open indoors with modifications under both tiers, while restaurants and gyms---which have larger estimated spillovers---were outdoor only under the purple tier and open indoors with modifications under the red tier (Table \ref{tab:blueprint_sector}).}

We also observe significant positive $RR$ effects in 22 POI groups (21 with Bonferroni correction), as in, visits increase significantly when both the CBG and POI are in red, compared to when they are both in purple. Furthermore, in most POI groups, the $PR$ effect is \textit{larger} than the $RR$ effect (although not always significantly so). This suggests an interaction effect: individuals not only spill over into adjacent counties when those counties become less restrictive, but also the spillover is larger if their home counties are more restrictive. 
Finally, we observe a varying effect of $RP$, which represents when the CBG changes from purple to red, while the POI remains in purple. The effect is slightly positive or negative for some POI groups, but significant in neither direction for most. We hypothesize that two mechanisms take place here: on one hand, since the POI is in a more restricted tier than the CBG, it becomes less appealing; on the other hand, since the CBG opened up, its population is more willing to travel. These counteracting mechanisms may explain the varying and weak $RP$ effects across POI groups. 

\paragraph{Local vs. global restrictions.}
To contrast local and global approaches to policymaking, we use our fitted model to compare counterfactual mobility reductions under county-level vs. hypothetical statewide restrictions.
Formally, let $\mathbf{T} \in \mathbb{R}^{58}$ represent the treatment vector for all counties. For each county $A$, we estimate this county's expected mobility (out-degree in the mobility network) under three treatment conditions: when the entire state is in the red tier ($\mathbf{T}_R$), when the entire state is in the purple tier ($\mathbf{T}_P$), and when only this county is in purple while the rest of the state remains in red ($\mathbf{T}_A$). 
We then compare the mobility reduction that a county would experience by going to purple on its own, relative to the statewide shutdown, where all counties go to purple:
\begin{align}
    r(A) &= \frac{\mathbb{E}[out(A) | \mathbf{T}_R] - \mathbb{E}[out(A) | \mathbf{T}_A]}{\mathbb{E}[out(A) | \mathbf{T}_R] - \mathbb{E}[out(A) | \mathbf{T}_P]}.
\end{align}
We calculate $\mathbb{E}[out(A) | \mathbf{T}]$ as the sum over within-county visits and out-of-county visits:
\begin{align}
    \mathbb{E}[out(A) | \mathbf{T}] =  \mathbb{E}[Y_{AA} | T_A] + \hspace{-2mm} \sum_{B \in N(A)} \mathbb{E}[Y_{AB} | T_A, T_B],
\end{align}
where $Y_{AB}$ represents the total number of visits from any CBG in county $A$ to any POI in county $B$. When we use our fitted model to compute the conditional expectation of $Y_{AB}$ given tiers, we assume $Z = 0$ for all CBGs and POIs, since our RD-based framework estimated tier effects at the joint cutoff. We also marginalize over the remaining dynamic covariate, the CBG's weekly device count, by taking each CBG's average device count \extended{over the 9-week period that we study. 
In the Appendix, we describe how to efficiently compute $Y_{AB}$ for all pairs of counties and possible tier pairs.}

We estimate that counties applying local restrictions can only achieve, on average, 
54.0\% (46.4\%--61.7\%) of the reduction in mobility that they would experience under a statewide shutdown. 
\extended{Small counties are particularly affected, keeping only 41.7\% (31.0\%--52.4\%) of their statewide mobility reduction, while large counties retain 62.1\% (56.3\%--67.9\%).}
While we assume that the reduction in mobility \textit{within} the county stays the same between local and global regimes, the difference arises from the increase in \textit{out-of-county visits} when all surrounding counties are less restricted in the red tier; \extended{this is why smaller counties are especially hard-hit, since a larger fraction of their mobility tends to be out-of-county}.
We also consider a less extreme case, where instead of having all surrounding counties in red, we use the actual configuration of tiers from the Blueprint. We still observe serious costs to efficacy in these more realistic settings: over the course of our study period, as the number of counties in purple fell from 40 to 11 to 3, the average percent of mobility reduction kept for counties in purple (compared to statewide shutdown) fell from 94\% to 75\% to 65\% \extended{(Figure \ref{fig:rwa_over_real_tiers})}.
These substantial decreases in efficacy demonstrate the cost of spillovers on local policies.

\paragraph{Balancing efficacy and flexibility.} 
\begin{figure}
    \centering
    \includegraphics[width=\linewidth]{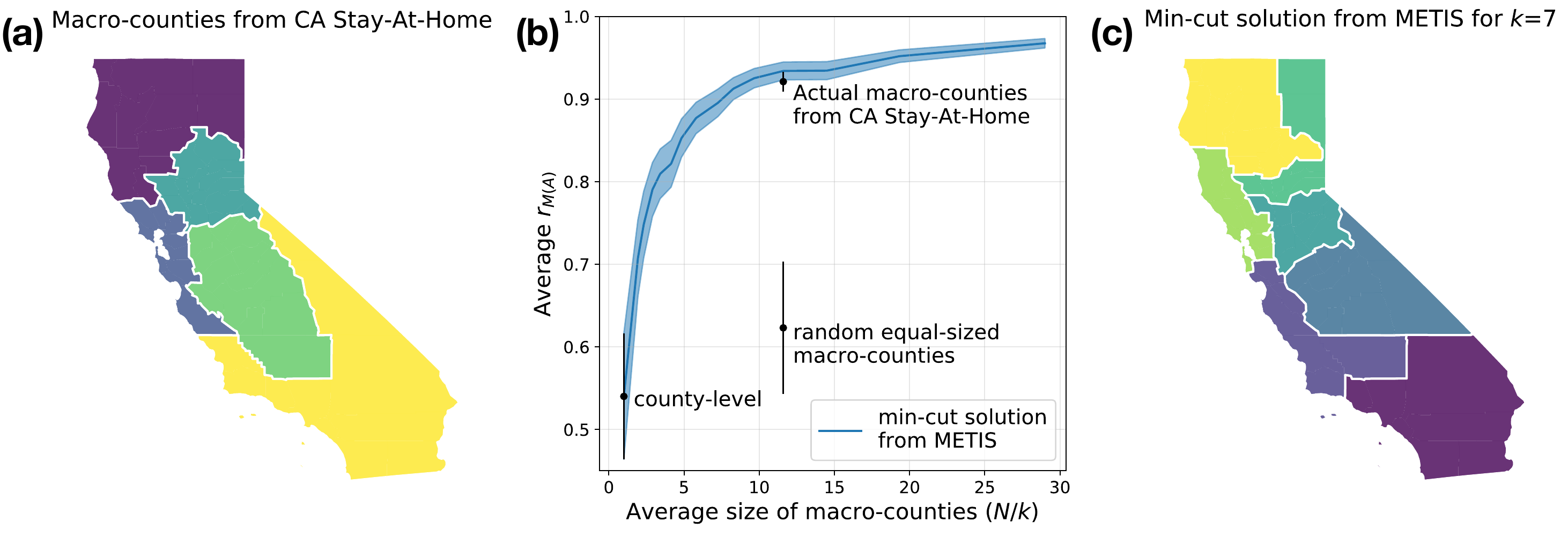}
    \caption{Macro-county restrictions. (a) Actual macro-counties from California's Stay-At-Home Order. (b) Trade-off between flexibility (lower macro-county size) versus efficacy (higher $r_M(A)$), with 95\% confidence intervals. (c) Macro-county partition for $k=7$, computed by METIS.}
    \label{fig:pareto_curve}
\end{figure}

Although local policies are less effective in the presence of spillovers, global policies are often too blunt and inflexible. In our final analysis, we explore this trade-off between efficacy and flexibility across policies at different spatial scales. Instead of being entirely local (county-level) or global (statewide), intermediate strategies could be implemented at the macro-county level. California in fact pursued such a strategy with its Regional Stay-At-Home Order \citep{newsom2020stayathome} that grouped counties into 5 macro-counties, each containing 11--13 counties (Figure \ref{fig:pareto_curve}a). 
Given a county partition $M$, we extend our analysis to compute $r_M(A)$, the ratio of mobility reduction that each county $A$ would experience if only its \textit{macro-county} went to purple, compared to statewide shutdown.
When we use the county partition from California's Regional Stay-At-Home Order, we find that macro-county restrictions can achieve 92.1\% (90.9\%--93.3\%) of statewide mobility reductions. In contrast, if we use a random partitioning of counties into equal-sized segments, such restrictions only reach 62.3\% (54.3\%--70.4\%, 95\% CI includes randomness in partitioning) of statewide reductions. Thus, policies of intermediate scale are promising in their ability to balance efficacy and flexibility, but achieving that balance relies on optimizing how macro-counties are defined.

Given a desired number of macro-counties $k$, we show in the Appendix that we can find the optimal county partition that maximizes the average $r_M(A)$ over counties by solving a minimum $k$-cut problem, which seeks to partition the nodes of an undirected graph into $k$ disjoint sets while minimizing the total weight of edges between nodes in different sets.
We define our undirected graph as one between counties, where the edge weight $w_{AB}$ between two adjacent counties $A$ and $B$ is
\begin{align}
    w_{AB} = &\frac{\mathbb{E}[Y_{AB} | P,R] - \mathbb{E}[Y_{AB} | P,P]}{\mathbb{E}[out(A) | \mathbf{T}_R] - \mathbb{E}[out(A) | \mathbf{T}_P]} + \\
    &\frac{\mathbb{E}[Y_{BA} | P,R] - \mathbb{E}[Y_{BA} | P,P]}{\mathbb{E}[out(B) | \mathbf{T}_R] - \mathbb{E}[out(B) | \mathbf{T}_P]} \nonumber.
\end{align}
To achieve evenly sized macro-counties, we impose an additional constraint (common in balanced graph partitioning) that each set is no larger than $1.05 \cdot \frac{N}{k}$, where $N=58$ is the total number of counties.
While this problem is NP-hard, we can approximate the solution using METIS \citep{karypis1997metis}.
In Figure \ref{fig:pareto_curve}b, we display our solutions over a range of $k$. Smaller macro-county sizes are preferred for flexibility (x-axis), while higher $r_M(A)$ represents better efficacy (y-axis). We observe a clear trade-off between the two objectives; however, even small macro-counties---when grouped intelligently---yield large improvements in efficacy over county-level restrictions. For example, by just increasing the average macro-county size to 8 (still $1/7^{th}$ the total number of counties), we reach over 90\% of the full efficacy of the much more drastic statewide shutdown (Figure \ref{fig:pareto_curve}c).
\section{Conclusion}
Geographic spillovers arise in many domains, but there are few opportunities to reliably estimate spillover effects.
In this work, we identify a novel setting that is uniquely suitable for spillover analysis, California's Blueprint for a Safer Economy, which defined a set of policies applied with the same deterministic thresholds across 58 counties.
We leverage these properties to develop a causal inference framework that allows us to make unconfounded estimates of spillover movement between counties and we observe significant spillovers in many POI groups. Finally, we evaluate the cost of spillovers on policies across spatial scales, analyzing the trade-off between efficacy and flexibility.

Our work is not without limitations. First, SafeGraph's data does not cover all POIs or populations uniformly. To mitigate this issue, we control for CBG weekly device count, only estimate effects for the largest POI categories, and drop categories such as elementary schools that have unreliable coverage from cell phone apps. 
Second, our causal inference framework may not entirely satisfy SUTVA, the assumption that a unit's outcome is only influenced by its own treatment. 
In this work, we attempt to better satisfy SUTVA by modeling the effect of pairwise policies on cross-county movement, instead of only modeling the effect of a single county's policies on its population's mobility, as prior work has done. 
However, future work should explore interference beyond pairs; for example, mobility from county $A$ to $B$ may depend not only on $A$ and $B$'s policies but also on the policies of $A$'s other neighbors.
We also hope that future work will dive deeper into the complex trade-offs of policymaking for interconnected regions. 
In this work, we explored efficacy and flexibility, but other dimensions should be considered, such as
equity in the context of certain regions bearing disproportionate risks and unequal resources (e.g., with resourced areas better able to handle spikes in COVID-19 cases).
\section*{Acknowledgements}
S. C. was supported in part by an NSF Graduate Research Fellowship and the Meta PhD Fellowship. The authors thank Emma Pierson, Martin Saveski, Hamed Nilforoshan, and anonymous reviewers for helpful comments and discussions.

\bibliography{main}

\appendix
\renewcommand\thefigure{A\arabic{figure}}   
\setcounter{figure}{0}   
\renewcommand\thetable{A\arabic{table}}   
\setcounter{table}{0}    
\renewcommand\thesection{A\arabic{section}}   

\section{Details on Data and Model Fitting}
\paragraph{Data and code availability.} The code to run our experiments and regenerate figures is available online.\footnote{\url{https://github.com/snap-stanford/covid-spillovers}} We also make our constructed $Z$ variables available, to facilitate future research that uses them in regression discontinuity designs to estimate the effects of California Blueprint tiers on spillovers and other outcomes of interest.

Documentation about the California Blueprint for a Safer Economy is provided by the California Department of Public Health (CDPH), such as how tiers were assigned \citep{cdph2021blueprint, cdph2021healtheq} and what the tier restrictions were for different sectors (Table \ref{tab:blueprint_sector}). CDPH has also archived historical tier assignments and COVID-19 metrics per county over the course of the Blueprint.\footnote{\url{https://www.cdph.ca.gov/Programs/CID/DCDC/Pages/COVID-19/CaliforniaBlueprintDataCharts.aspx}}
Our mobility data comes from SafeGraph Weekly Patterns\footnote{\url{https://docs.safegraph.com/docs/weekly-patterns}}, which is available to researchers through Dewey.\footnote{\url{https://www.deweydata.io/}}
SafeGraph also provides each POI's ``top'' category (first 4 digits of NAICS code) and ``sub'' category (first 6 digits), which we use to learn heterogeneous effects for different POI groups (Table \ref{tab:poi-groups}).
Finally, we use data from the US Census Bureau's 5-year American Community Survey about census block groups, which is available online.\footnote{\url{https://www.census.gov/programs-surveys/acs/data.html}}

\begin{table*}
    \centering
    \begin{tabular}{|p{5cm}|p{5cm}|p{5cm}|}
        \hline 
          \textbf{Sector} & \textbf{Purple Tier} & \textbf{Red Tier} \\
         \hline 
          Critical Infrastructure (e.g., hospitals, emergency services, grocery stores, gas stations) & Open with modifications & Open with modifications \\
          \hline
          Limited Services (provides services with limited contact, e.g, laundry services, auto repair shops, pet grooming) & Open with modifications & Open with modifications \\
          \hline 
          Outdoor playgrounds \& recreational facilities & Open with modifications & Open with modifications \\
          \hline 
          Hotels and Lodging & Open with modifications & Open with modifications \\
          \hline 
          Hair salons \& barbershops & Open indoors with modifications & Open indoors with modifications \\
          \hline
          Personal Care Services & Open indoors with modifications & Open indoors with modifications \\
          \hline
          All Retail & Open indoors with modifications, max 25\% capacity & Open indoors with modifications, max 50\% capacity \\
          \hline 
          Shopping Centers (Malls, Destination Centers, Swap Meets) & Open indoors with modifications, max 25\% capacity & Open indoors with modifications, max 50\% capacity \\
          \hline 
          Museums, Zoos, and Aquariums & Outdoor only with modifications & Open indoors with modifications, max 25\% capacity \\
          \hline 
          Places of Worship & Outdoor encouraged; indoors allowed with modifications, max 25\% capacity & Open indoors with modifications, max 25\% capacity \\
          \hline 
          Movie Theaters & Outdoor only with modifications & Open indoors with modifications, max 25\% capacity or 100 people (whichever is fewer) \\
          \hline 
          Restaurants & Outdoor only with modifications & Open indoors with modifications, max 25\% capacity or 100 people (whichever is fewer) \\
          \hline
          Gyms and Fitness Centers & Outdoor only with modifications & Open indoors with modifications, max 10\% capacity \\
          \hline 
    \end{tabular}
    \caption{The California Blueprint for a Safer Economy's section-specific restrictions for the purple and red tiers. For the full list of tiers and sectors, see \citet{cdph2021blueprint}, ``Risk Criteria''.}
    \label{tab:blueprint_sector}
\end{table*}

\begin{table*}
    \centering
    \begin{tabular}{|p{2cm}|p{4cm}|p{7cm}|p{1.5cm}|}
        \hline 
          \textbf{NAICS code} & \textbf{Full Name} & \textbf{Description} & \textbf{\# POIs} \\
         \hline 
        4411 & Automobile Dealers & New car and old car dealers & 1429 \\
        \hline
        4413 & Automotive Parts, Accessories, and Tire Stores & Retailers for automotive parts and repair & 2034 \\
        \hline
        4441 & Building Material and Supplies Dealers & Retailers for home improvement goods, paint and wallpaper, tools and builders' hardware & 1146 \\
        \hline 
        4451 & Grocery Stores & Supermarkets, convenience retailers, vending machine operators & 6449 \\
        \hline 
        4551 & Department Stores & Department stores for apparel, jewelry, home furnishings, toys, etc. & 1138 \\
        \hline 
        4552 & General Merchandise Stores, including Warehouse Clubs and Supercenters & Warehouse clubs, supercenters, dollar stores, home and auto supply stores, variety stores & 2280 \\
        \hline
        4561 & Health and Personal Care Stores & Retailers for drugs (i.e., pharmacies), beauty supplies, optical goods, food supplements & 4227 \\
        \hline 
        4571 & Gasoline Stations & Gasoline stations, sometimes with convenience stores & 7514 \\
        \hline
        4581 & Clothing Stores & Sells clothing, clothing accessories (e.g., hats, gloves, wigs) & 1656 \\
        \hline
        4591 & Sporting Goods, Hobby, and Musical Instrument Stores & Retailers for sporting goods, hobbies, toys, games, sewing and needlework supplies, musical instruments and supplies & 3362 \\
        \hline 
        4594 & Office Supplies, Stationery, and Gift Stores & Retailers for office supplies, stationery, office equipment, greeting cards, decorations & 1212 \\
        \hline 
        4595 & Used Merchandise Stores & Sells used goods, antiques, auctions & 1292 \\
        \hline 
        4599 & Other Miscellaneous Store Retailers & Retailers for pet supplies, art dealers, mobile home dealers, smoking supplies, other miscellaneous things & 3697 \\
        \hline
        5311 & Lessors of Real Estate & Lessors of real-estate for residential, non-residential (e.g., malls), and storage purposes & 3144 \\
        \hline
        7121 & Museums, Historical Sites, and Similar Institutions & Museums, historical sites, zoos, gardens, and nature parks & 7511 \\
        \hline
        713940 & Fitness and Recreational Sports Centers & Sports facilities, including exercise centers, ice or roller skating rinks, tennis club facilities, and swimming pools & 4730 \\
        \hline 
        7139 & Other Amusement and Recreation Industries & Golf courses, skiing facilities, marinas, bowling centers & 1834 \\
        \hline 
        7211 & Traveler Accommodation & Hotels, motels, casino hotels, bed-and-breakfasts & 2252 \\
        \hline 
        7224 & Drinking Places (Alcoholic Beverages) & Bars, taverns, nightclubs & 1791 \\
        \hline
        722511 & Full-Service Restaurants & Provides food services to patrons who order and are served while seated, then pay after & 21972 \\
        \hline
        722513 & Limited-Service Restaurants & Provides food services where patrons order and pay before eating; food may be consumed on premises, taken out, or delivered & 15074 \\
        \hline
        722515 & Snack and Nonalcoholic Beverage Bars & Prepares specialty snacks (e.g., ice cream) or non-alcoholic beverages (e.g., coffee) & 8528 \\
        \hline
        8131 & Religious Organizations & Churches, religious temples, synagogues, mosques, monasteries & 1065 \\
        \hline 
        -- & Other & All other POIs that were not in one of these groups & 20181 \\
        \hline
    \end{tabular}
    \caption{POI groups for which we learn heterogeneous treatment effects. We keep all POIs in California with at least 30 non-zero visits to CBGs, which leaves 128,655 POIs. Then, as POI groups, we keep all `top''-categories (first 4 digits of the NAICS code) with at least 1000 POIs (besides ``Elementary and Secondary Schools'', which we drop due to poor coverage from cell phone data) and 4 ``sub''-categories (first 6 digits of NAICS code) of interest. The category descriptions are based on \url{https://www.naics.com/six-digit-naics/}. }
    \label{tab:poi-groups}
\end{table*}

\begin{table}
    \centering
    \begin{tabular}{|p{1.8cm}|p{1.5cm}|p{1.5cm}|p{2cm}|}
        \hline 
          Week & \# Counties in Purple & \# Counties in Red & \# Data Points \\
         \hline 
          2021-02-01 & 9 & 0 & 2,659,022 \\
          2021-02-08 & 14 & 1 & 26,031,894 \\
          2021-02-15 & 20 & 2 & 58,585,151 \\
          2021-02-22 & 20 & 9 & 64,334,107 \\
          2021-03-01 & 28 & 16 & 497,292,295 \\
          2021-03-15 & 10 & 37 & 506,971,688 \\
          2021-03-22 & 7 & 30 & 204,963,737 \\
          2021-03-29 & 3 & 31 & 47,479,762 \\
        \hline 
        \textbf{Total} & \textbf{111} & \textbf{126} & \textbf{1,408,317,656}\\
        \hline
    \end{tabular}
    \caption{Size of the data we keep for fitting our model. We keep counties that are in the purple or red tier, comply with the expected assignment based on $Z$ (its tier is red if and only if $Z < 0$), and its $Z$ variable lies within a bandwidth $h=5$ of 0. Then, we keep all CBG-POI data points between adjacent kept counties. We also drop the week of March 8, 2021, since the purple-red threshold for adjusted case rate was changed in the middle of the week, due to the statewide vaccine goal being met.}
    \label{tab:data_size}
\end{table}

\begin{table*}
    \centering
    \begin{tabular}{|l|l|l|l|l|}
        \hline 
          & $PP$ & $PR$ & $RP$ & $RR$ \\
         \hline 
        $N$ & 126 & 71 & 74 & 177 \\
        \hline 
        1 & 6067 $\rightarrow$ 6061, 7.6\% & 6077 $\rightarrow$ 6067, 8.9\% & 6067 $\rightarrow$ 6077, 10.5\% & 6067 $\rightarrow$ 6061, 7.6\%  \\
        2 & 6037 $\rightarrow$ 6059, 5.8\% & 6067 $\rightarrow$ 6113, 8.3\% & 6113 $\rightarrow$ 6067, 8.3\% & 6065 $\rightarrow$ 6071, 7.5\% \\
        3 & 6013 $\rightarrow$ 6001, 5.1\% & 6075 $\rightarrow$ 6081, 6.4\% & 6017 $\rightarrow$ 6067, 8.1\% & 6037 $\rightarrow$ 6059, 6.0\% \\
        4 & 6059 $\rightarrow$ 6037, 4.0\% & 6029 $\rightarrow$ 6037, 5.9\% & 6107 $\rightarrow$ 6019, 7.5\% & 6071 $\rightarrow$ 6065, 5.6\% \\
        5 & 6061 $\rightarrow$ 6067, 3.7\% & 6115 $\rightarrow$ 6101, 5.9\% & 6081 $\rightarrow$ 6075, 5.3\% & 6065 $\rightarrow$ 6059, 4.9\% \\
        \hline
    \end{tabular}
    \caption{Distribution of top 5 most-represented adjacent county pairs per treatment condition. $N$ represents the number of unique county pairs seen per treatment condition.}
    \label{tab:h5_data_stats}
\end{table*}

\paragraph{Bandwidth selection.}
As we describe in Section \ref{sec:causal}, we filter the data based on a number of criteria, including that $Z_{iw}$ for CBG $c_i$'s county and $Z_{jw}$ for POI $p_j$'s county both lie within a bandwidth $h$ of 0 (since $Z=0$ is the threshold between assignment to the purple tier and red tier). 
Bandwidth selection introduces a bias-variance trade-off, with larger bandwidths corresponding to greater bias but reduced variance. We err on the side of larger bandwidths, so that we retain enough representation from different county pairs for each of the treatment conditions. When we use $h=5$ in our experiments and apply all our other filtering criteria, we are left with 1,408,317,656 data points overall (Table \ref{tab:data_size}).

684,604 of those data points represent non-zero visits, and among those, 289,199 belong to the $PP$ treatment condition, 48,608 to $PR$, 37,243 to $RP$, and 309,554 to $RR$. 
In Table \ref{tab:h5_data_stats}, we list, for each treatment condition, the number of unique adjacent county pairs that appear for this condition, the top 5 most-represented pairs, and the proportion of all non-zero data points that each pair accounts for within this treatment condition. We see that all treatment conditions, including the less common $PR$ and $RP$, still retain substantial diversity across counties, with over 70 unique pairs for each condition and no single county or county pair seriously dominating the data. The county pairs that appear more often are, as expected, the ones with a large number of CBGs in the source county and a large number of POIs in the target county. 

\begin{figure}
    \centering
    \begin{subfigure}[b]{0.47\textwidth}
       \includegraphics[width=\linewidth]{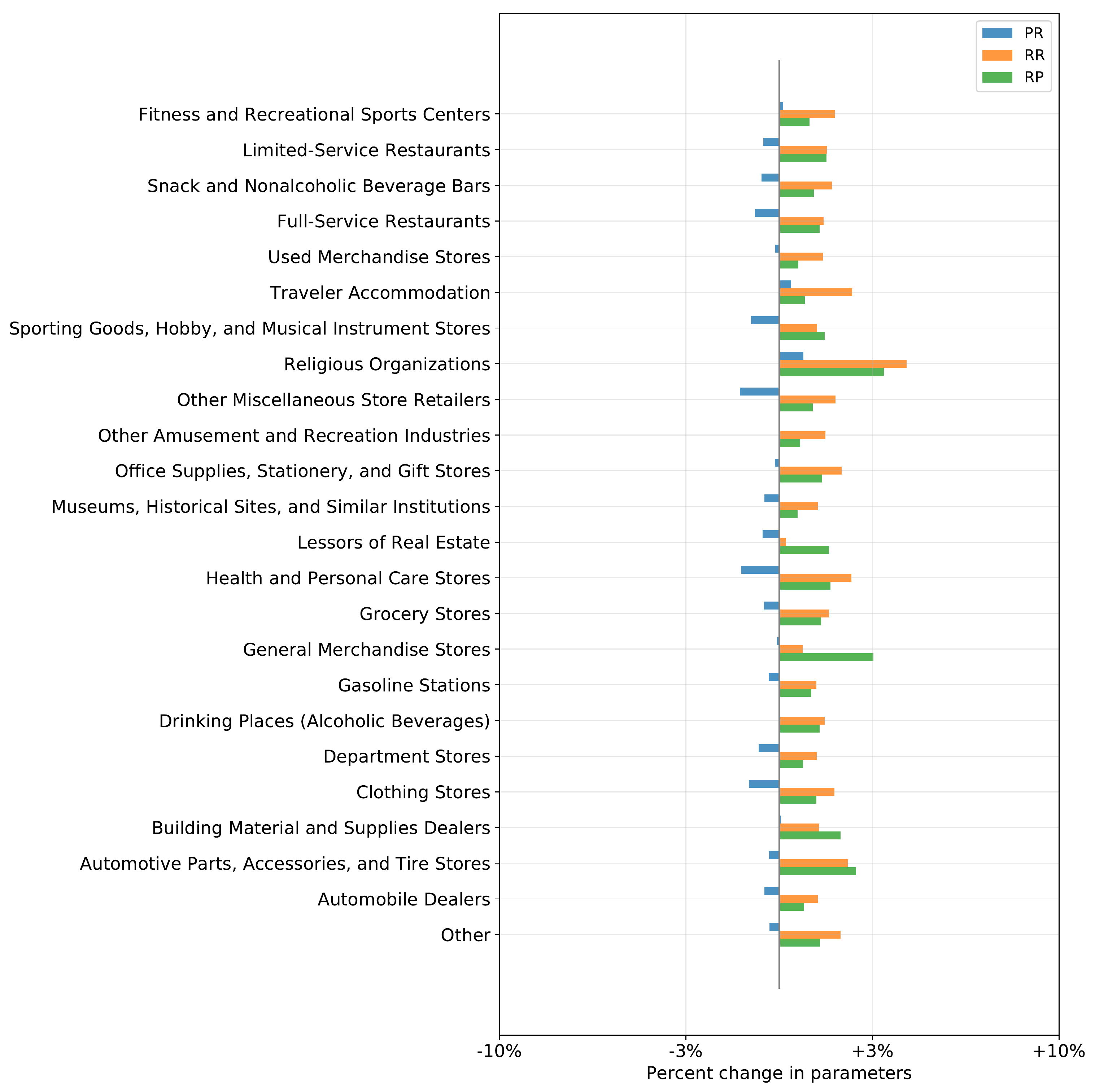}
       \caption{$h=4$ vs. $h=5$.}
       \label{fig:h4_v_h5} 
    \end{subfigure}
    
    \begin{subfigure}[b]{0.47\textwidth}
       \includegraphics[width=\linewidth]{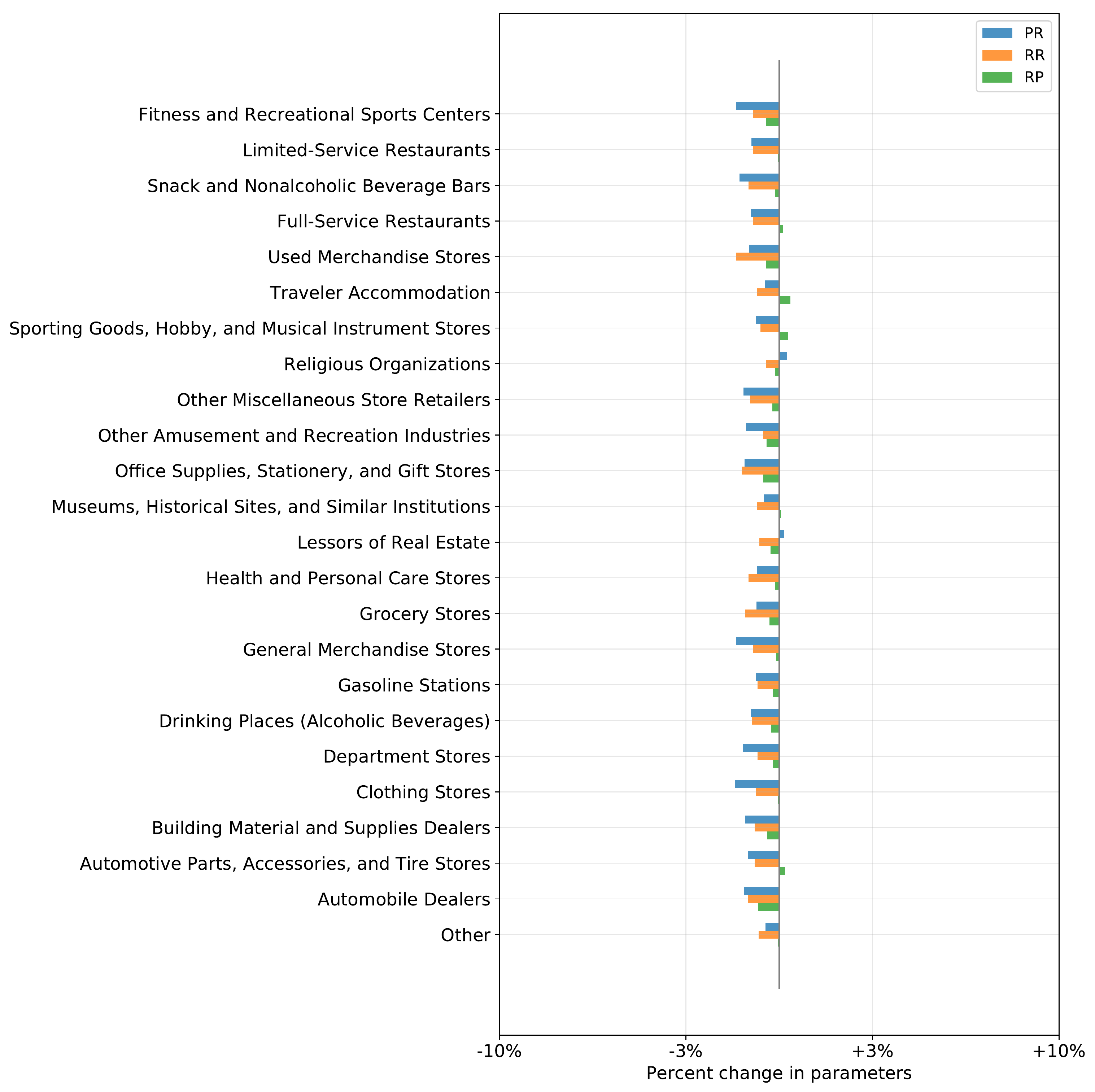}
       \caption{$h=6$ vs. $h=5$.}
       \label{fig:h6_v_h5}
    \end{subfigure}
    \caption{Percent change in estimated spillover parameters for different choices of bandwidth $h$.}
\label{fig:bandwidth_results}
\end{figure}

We also conduct sensitivity analyses with $h=4$ and $h=6$. Compared to the estimated parameters when $h=5$, the estimated parameters when $h=4$ typically only change by 2-3\% and at most 5\% (Figure \ref{fig:h4_v_h5}). The change is even smaller when we compare $h=6$ to $h=5$; the change is at most 2\% and mostly smaller than 1\% (Figure \ref{fig:h6_v_h5}). 

\paragraph{Loss-corrected negative sampling.} 
We perform negative sampling such that we sample each zero data point $(i, j, w)$ with probability $s_{ijw}$. Then, we fit our model on sample $S$, which contains all of the non-zero data points from the original data set and our sampled zero data points. However, negative sampling biases our model parameters by greatly reducing the number of zeros in the training data. To correct this bias, we weight each sampled data point by $\frac{1}{s_{ijw}}$ when computing the overall loss (negative log likelihood). These corrections ensure that our stochastic gradient $\nabla_\theta \mathcal{L}_{S}(\theta)$, computed over sample $S$, forms an unbiased estimate of the true gradient $\nabla_\theta \mathcal{L}(\theta)$, computed over the full data.

The proof of this is very straightforward, utilizing the fact that the negative log likelihood is a sum over negative log likelihoods per data point, and then applying linearity of expectation. Let indicator variable $b_{ijw} \sim \textrm{Bern}(s_{ijw})$ represent whether data point $(i, j, w)$ is in our sample $S$:
\begin{align}
    \nabla_\theta \mathcal{L}_{S}(\theta) &= -\sum_{i, j, w \in C} b_{ijw} \frac{1}{s_{ijw}} \nabla_\theta \ln(\Pr(Y_{ijw} | \theta)) \\
    \mathbb{E}[\nabla_\theta \mathcal{L}_{S}(\theta)] &= -\sum_{i, j, w \in C} \mathbb{E}[b_{ijw}] \frac{1}{s_{ijw}} \nabla_\theta \ln(\Pr(Y_{ijw} | \theta)) \\
    &= -\sum_{i, j, w \in C} \nabla_\theta \ln(\Pr(Y_{ijw} | \theta)) = \nabla_\theta \mathcal{L}(\theta).
\end{align}
Thus, incorporating a correction $\frac{1}{s_{ijw}}$ into the loss per data point ensures that the stochastic gradient forms an unbiased estimate of the true gradient, \textit{regardless of the negative sampling scheme used}. 

However, different negative sampling schemes, i.e., different choices of $s_{ijw}$, may be preferable in order to reduce the variance of the stochastic gradient. First, note that the variance of the stochastic gradient is the sum of the variances per data point, since each indicator variable $b_{ijw}$ is sampled independently. Second, observe that the non-zero data points contribute no variance, since they are sampled with probability 1. So, the variance of the stochastic gradient is a sum of variances over the zero data points, which we refer to as $C_0$:
\begin{align}
    &\textrm{Var}[\nabla_\theta \mathcal{L}_{S}(\theta)] = \sum_{i, j, w \in C_0} \textrm{Var}[b_{ijw} \frac{1}{s_{ijw}} \nabla_\theta \ln(\Pr(Y_{ijw} = 0 | \theta))] \\
    &= \sum_{i, j, w \in C_0} s_{ijw}(1-s_{ijw}) (\frac{1}{s_{ijw}} \nabla_\theta \ln(\Pr(Y_{ijw} = 0 | \theta)))^2 \\
    &= \sum_{i, j, w \in C_0} (\frac{1}{s_{ijw}} - 1) (\nabla_\theta \ln(\Pr(Y_{ijw} = 0 | \theta)))^2.
\end{align}
Since $s_{ijw}$ is a probability, then $\frac{1}{s_{ijw}} \geq 1$. We can see, firstly, that larger sampling probabilities reduce the variance, and when $s_{ijw} = 1$ for all $i, j, w$ (meaning we sample all zero data points with probability 1), the variance is 0. Barring an increase in sample size, observe that we would want to prioritize sampling ``harder'' data points, i.e., those with larger gradients, where the model is learning more from $Y_{ijw} = 0$. While we cannot know before sampling which data points would have larger gradients, we can use proxies, such as distance between the CBG and POI, to estimate harder samples. Thus, we set $s_{ijw} \propto \frac{1}{1 + d_{ij}}$, where $d_{ij}$ is the distance between CBG $c_i$ and POI $p_j$. Then, we scale the $s_{ijw}$ terms such that in expectation, we sample a certain percentage, such as 2\%, of the zero data points. 

\begin{figure}
    \centering
    \includegraphics[width=\linewidth]{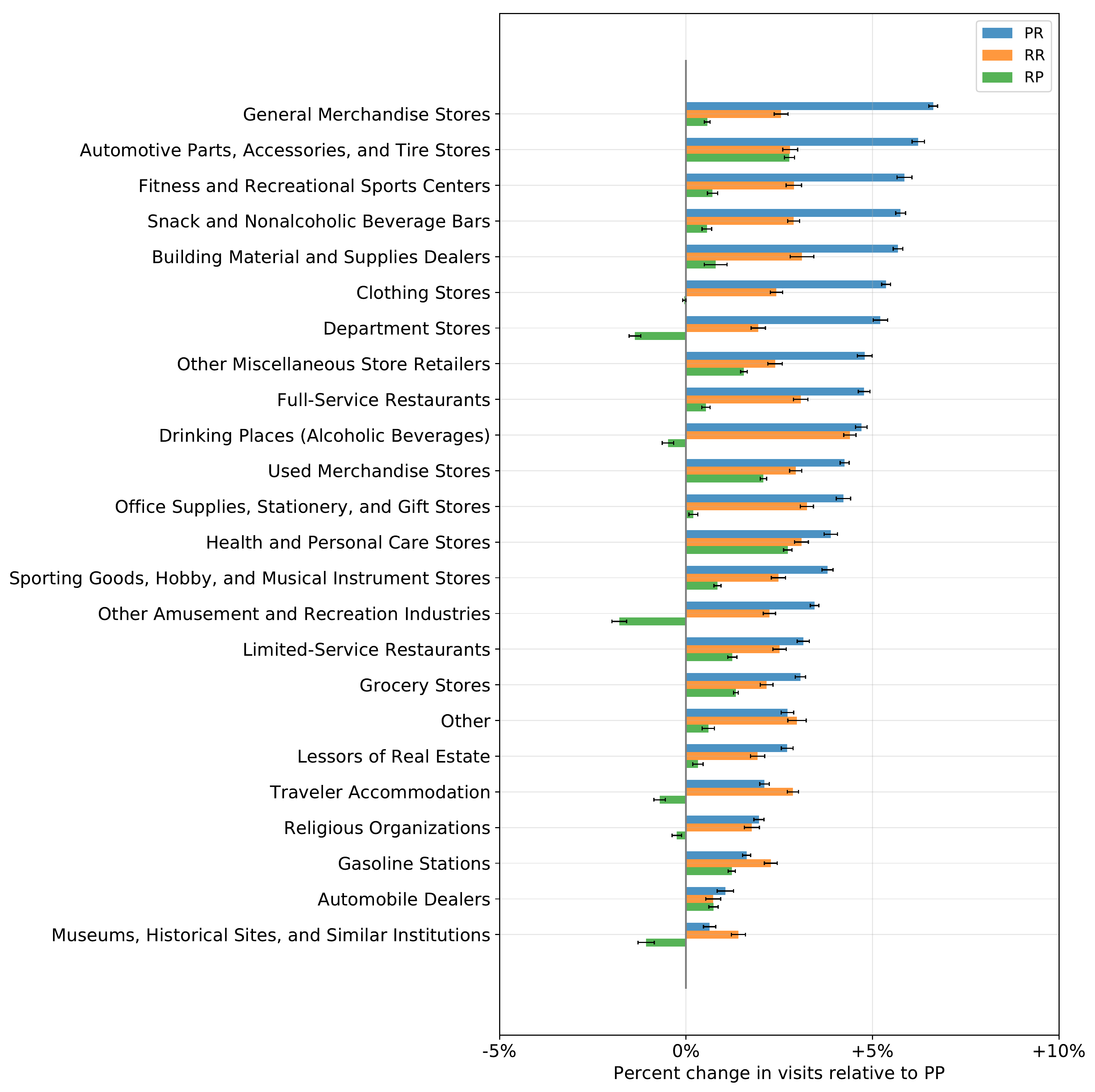}
    \caption{Estimated spillover effects across POI groups, with 95\% confidence intervals only capturing uncertainty over negative samples and not over the data.}
    \label{fig:neg_sample_cis}
\end{figure}

Recall that in our bootstrapping procedure, we sample the non-zero data points with replacement and draw a fresh negative sample in every trial. This procedure allows us to capture uncertainty in the data as well as uncertainty from negative sampling.
We conduct an additional experiment where we only capture uncertainty from negative sampling, by conducting 10 more trials with fresh negative samples but without sampling the non-zero data points. We use the same negative sampling scheme as we do in our main experiments, sampling 2\% of zero data points with distance-weighted sampling probabilities. Compared to our main results (Figure \ref{fig:spillover_results}), we show in Figure \ref{fig:neg_sample_cis} that negative sampling only accounts for a very small proportion of the overall uncertainty, confirming that our negative sampling scheme is sufficient.

\paragraph{Model fitting.} 
\begin{figure}
    \centering
    \begin{subfigure}[b]{0.47\textwidth}
       \includegraphics[width=\linewidth]{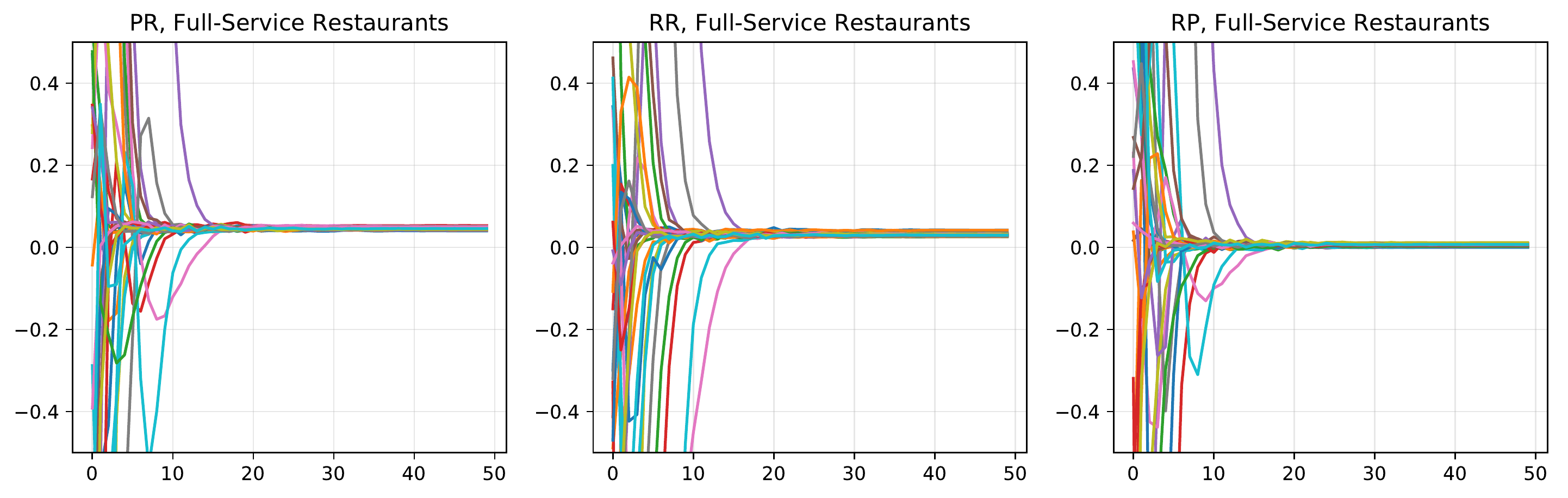}
       \caption{Full-Service Restaurants.}
    \end{subfigure}
    
    \begin{subfigure}[b]{0.47\textwidth}
       \includegraphics[width=\linewidth]{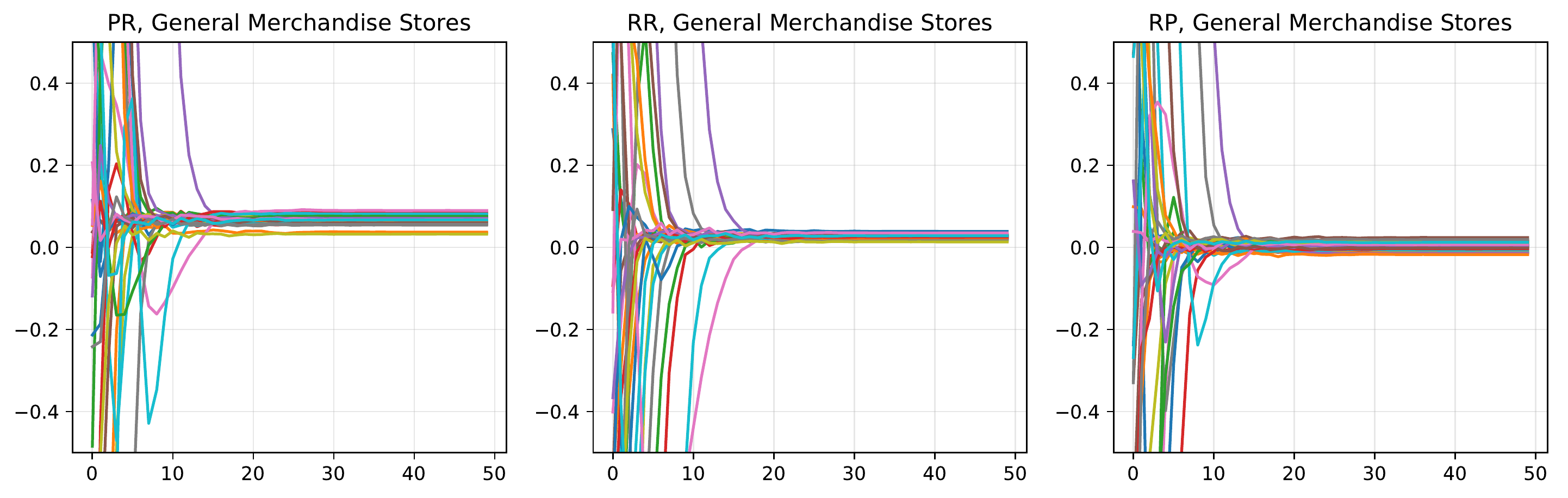}
       \caption{General Merchandise Stores.}
    \end{subfigure}
    \caption{Model parameters converge over epochs. We plot one line for each of the 30 trials. The example model parameters are the spillover weights $\beta_{PR}$, $\beta_{RR}$, and $\beta_{RP}$ for two POI groups.}
\label{fig:convergence}
\end{figure}

As described in Section \ref{sec:causal}, we quantify uncertainty with 30 trials.
In each trial, we re-sample the data and fit the model using loss-corrected gradient descent, running for 50 epochs.
Empirically, the model parameters reliably converge within this number of epochs and, across trials, parameters also converge to similar values (Figure \ref{fig:convergence}). 

\section{Details For Computing Results}
\paragraph{Local vs. global restrictions.} 
In this section, we discuss in more detail how we conduct our analysis of county-level vs. hypothetical statewide restrictions. First, let $\mathbf{T} \in \mathbb{R}^{58}$ represent a treatment vector for all 58 counties. Then, let $\mathbf{T}_R$ represent the all-red treatment condition, $\mathbf{T}_P$ represent the all-purple treatment condition, and $\mathbf{T}_A$ represent the condition where all counties are in red except county $A$, which is in purple. Recall that our goal is to compare the reduction in county $A$'s expected out-degree, $\mathbb{E}[out(A)]$, from $\mathbf{T}_R$ to $\mathbf{T}_A$, compared to $\mathbf{T}_P$:
\begin{align}
    r(A) &= \frac{\mathbb{E}[out(A) | \mathbf{T}_R] - \mathbb{E}[out(A) | \mathbf{T}_A]}{\mathbb{E}[out(A) | \mathbf{T}_R] - \mathbb{E}[out(A) | \mathbf{T}_P]}.
\end{align}

We calculate $\mathbb{E}[out(A) | \mathbf{T}]$ as the sum over within-county visits and out-of-county visits:
\begin{align}
    \mathbb{E}[out(A) | \mathbf{T}] = \mathbb{E}[Y_{AA} | T_A] + \sum_{B \in N(A)} \mathbb{E}[Y_{AB} | T_A, T_B],
\end{align}
where $Y_{AB}$ represents the total number of visits from any CBG in $A$ to any POI in $B$. When we use our fitted model to compute the conditional expectation of $Y_{AB}$ given tiers, we assume $Z = 0$ for all CBGs and POIs, since our RD-based framework estimated tier effects at the joint cutoff. We also marginalize over the remaining dynamic covariate, the CBG's weekly device count, by taking each CBG's average device count over the 9-week period that we study. This produces a static vector $\mathbf{X}_{ij}$, representing the CBG and POI covariates.

Using our zero-inflated Poisson regression model, we compute $\mathbb{E}[Y_{AB}]$ as the sum over visits from each CBG in $A$ to each POI in $B$, where we use POI group-specific weights $\beta_{g_j, T_A, T_B}$ to capture heterogeneous tier effects:
\begin{align}
    &\mathbb{E}[Y_{AB} | T_A, T_B] = \sum_{i \in A} \sum_{j \in B} \mathbb{E}[Y_{ij} | T_A, T_B] \\
    &= \sum_{i \in A} \sum_{j \in B} \frac{1}{1 + \alpha_1 d_{ij}^{\alpha_2}} \exp(\beta_0 + \mathbf{\beta}_3^T\mathbf{X}_{ij} + \beta_{g_j, T_A, T_B}) \\
    &= \sum_{g} \exp(\beta_{g, T_A, T_B}) \underbrace{\sum_{i} \sum_{j; g_j = g} \frac{1}{1 + \alpha_1 d_{ij}^{\alpha_2}} \exp(\beta_0 + \mathbf{\beta}_3^T\mathbf{X}_{ij})}_{\phi(g, A, B)}.
\end{align}
We can simplify computation by pre-computing weights $\phi(g, A, B)$ for each adjacent county pair and POI group $g$. Then, the numerator of $r(A)$ becomes
\begin{align}
    \mathbb{E}[&out(A) | \mathbf{T}_R] - \mathbb{E}[out(A) | \mathbf{T}_A] \\
    &= \mathbb{E}[Y_{AA} | R] - \mathbb{E}[Y_{AA} | P] + \\
    &\sum_{B \in N(A)} \mathbb{E}[Y_{AB} | R, R] - \mathbb{E}[Y_{AB} | P, R] \nonumber \\
    &= \mathbb{E}[Y_{AA} | R] - \mathbb{E}[Y_{AA} | P] + \\
    &\sum_{g \in G} \exp(\beta_{g, R, R}) - \exp(\beta_{g, P, R}) \sum_{B \in N(A)} \phi(g, A, B). \nonumber
\end{align}
Similarly, the denominator is 
\begin{align}
    \mathbb{E}[&out(A) | \mathbf{T}_R] - \mathbb{E}[out(A) | \mathbf{T}_P] \\
    &= \mathbb{E}[Y_{AA} | R] - \mathbb{E}[Y_{AA} | P] + \\
    &\sum_{g \in G} \exp(\beta_{g, R, R}) - \exp(\beta_{g, P, P}) \sum_{B \in N(A)} \phi(g, A, B). \nonumber
\end{align}
We fit a separate model on only within-county visits to calculate $\mathbb{E}[Y_{AA}]$ and use it to estimate the change in within-county visits from the red to purple tier.
With this analysis, we estimate that counties applying local restrictions can only achieve, on average, 54.0\% (46.4\%--61.7\%) of the reduction in mobility that they would experience under a statewide shutdown. 
We also compute the averages over only the 23 small counties (population under 106,000), which yields 41.7\% (31.0\%--52.4\%), and over the 35 large counties, which yields 62.1\% (56.3\%--67.9\%).

We also analyze a less extreme setting, where instead of $\mathbf{T}_A$, where all counties are in red except county $A$, we consider the actual tier configuration from the California Blueprint. First, we take the real assignments from a week $w$ in our study period, such as March 15, 2021, when 11 counties were in the purple tier, 42 were in the red tier, 4 were in the orange tier, and 1 was in the yellow tier (Figure \ref{fig:data_plots}a). We construct a new treatment vector $\mathbf{T}_w^{'}$, where a county is assigned to the purple tier if it was in the purple tier in week $w$ and assigned to the red tier, otherwise. Then, we compute a similar ratio $r_w(A)$:
\begin{align}
    r_w(A) &= \frac{\mathbb{E}[out(A) | \mathbf{T}_R] - \mathbb{E}[out(A) | \mathbf{T}_w^{'}]}{\mathbb{E}[out(A) | \mathbf{T}_R] - \mathbb{E}[out(A) | \mathbf{T}_P]}.
\end{align}
$r_w(A)$ represents the proportion of mobility reduction kept under this more realistic scenario, compared (as before) to a statewide shutdown. Then, we take the average $r_w(A)$ over the counties that were assigned to purple in week $w$. 

\begin{figure}
    \centering
    \includegraphics[width=0.8\linewidth]{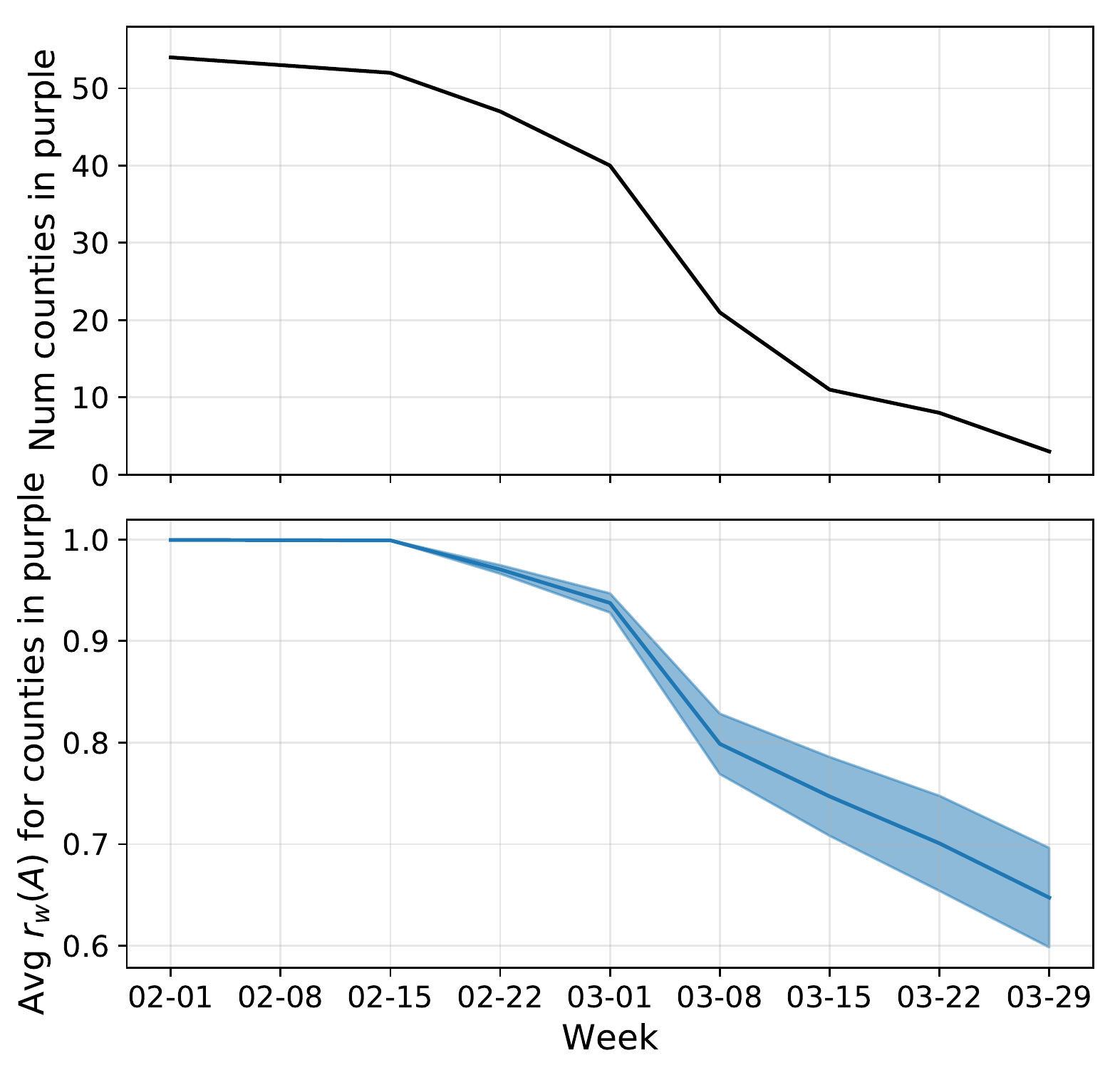}
    \caption{Evaluating the efficacy of realistic tier configurations. Over the course of our 9-week study period, we visualize the number of counties kept in the purple tier based on the real Blueprint tier assignments from that week (top) and the average percentage of mobility reduction kept for those counties in purple, with 95\% confidence intervals (bottom).}
    \label{fig:rwa_over_real_tiers}
\end{figure}

Over the course of our study period, the number of counties in purple fell from 54 to 3. As expected, as the number of counties in purple fell, the amount of mobility reduction retained for the counties still in purple fell as well (Figure \ref{fig:rwa_over_real_tiers}). 
For example, by the week of March 15, 2021, when there were only 11 counties left in the purple tier, those counties only kept 74.7\% (70.8\%--78.6\%) of the statewide mobility reduction. 
Two weeks later, the remaining 3 counties in the purple tier could only retain 64.7\% (59.8\%-69.6\%) of their statewide reductions. 
While these percentages are higher than the worst-case (54\%, if each county goes to purple alone), they are still far below the full efficacy of the statewide restrictions, demonstrating the cost of spillovers on local policy regimes even under more realistic realizations of policies.

\paragraph{Balancing efficacy and flexibility.}
To analyze macro-county restrictions, we introduce $\mathbf{T}_{M(A)}$, which represents the treatment condition where all counties are in red except county $A$'s \textit{macro-county}, which is in purple. Then, we define $r_M(A)$ for a county partition $M$ as a a simple extension of $r(A)$:
\begin{align}
     r_M(A) = \frac{\mathbb{E}[out(A) | \mathbf{T}_R] - \mathbb{E}[out(A) | \mathbf{T}_{M(A)}]}{\mathbb{E}[out(A) | \mathbf{T}_R] - \mathbb{E}[out(A) | \mathbf{T}_P]}
\end{align}
We can also compute this quantity efficiently using the pre-computed weights per county pair and POI group.

To find optimal county partitions, we define an undirected graph $G$ between counties, where the edge weight $w_{AB}$ between two adjacent counties $A$ and $B$ is
\begin{align}
    w_{AB} = &\frac{\mathbb{E}[Y_{AB} | P,R] - \mathbb{E}[Y_{AB} | P,P]}{\mathbb{E}[out(A) | \mathbf{T}_R] - \mathbb{E}[out(A) | \mathbf{T}_P]} + \\
    &\frac{\mathbb{E}[Y_{BA} | P,R] - \mathbb{E}[Y_{BA} | P,P]}{\mathbb{E}[out(B) | \mathbf{T}_R] - \mathbb{E}[out(B) | \mathbf{T}_P]} \nonumber,
\end{align}
and the edge weight between non-adjacent counties is 0.
Now, we show that finding the county partition that maximizes the average $r_M(A)$, for a fixed number of macro-counties $k$, is equivalent to solving a minimum $k$-cut problem on $G$.
First, observe which parts of $r_M(A)$ actually vary with our choice of $M$. The denominator is constant and the numerator we can expand out to
\begin{align}
    &\mathbb{E}[Y_{AA}|R] - \mathbb{E}[Y_{AA}|P] + \sum_{\substack{B \in N(A);\\M(A) = M(B)}} \mathbb{E}[Y_{AB} | R,R] - \mathbb{E}[Y_{AB} | P, P]\\
     &+ \sum_{\substack{B \in N(A);\\M(A) \neq M(B)}} \mathbb{E}[Y_{AB} | R,R] - \mathbb{E}[Y_{AB} | P, R].  \nonumber
\end{align}
The terms for within-county visits, $Y_{AA}$, are also constant, and the remaining terms are the summations over neighbors of $A$. At best, all of $A$'s neighbors are in its macro-county, so $M(A)=M(B)$ applies to all neighbors. If we consider moving one neighbor $B$ outside of $M(A)$, this will add $c_{BA}$ to $r_M(A)$:
\begin{align}
    c_{BA} = \frac{\mathbb{E}[Y_{AB}|P, P]) - \mathbb{E}[Y_{AB}|P, R])}{\mathbb{E}[out(A) | \mathbf{T}_R] - \mathbb{E}[out(A) | \mathbf{T}_P]}.
\end{align}
This quantity tends to be negative since we showed that $\mathbb{E}[Y_{AB}|P, R]$ is typically larger than $\mathbb{E}[Y_{AB}|P, P]$, due to spillovers. Furthermore, the more spillover there is from $A$ to $B$, the more negative this quantity will be. Thus, to maximize $r_M(A)$, we want to choose a partition $M$ that maximizes $c_{BA}$ and $c_{AB}$ (i.e., minimizes spillovers) over all pairs of adjacent counties that are \textit{not} in the same macro-county. 
This is equivalent to finding the minimum $k$-cut in $G$, since we defined the edge weight $w_{AB}$ as the sum of $-c_{BA}$ and $-c_{AB}$.
In practice, we use the mean model parameters over our 30 trials to construct $G$. Then, to approximate a solution to the minimum $k$-cut problem, we run the METIS algorithm\footnote{\url{https://metis.readthedocs.io/en/latest/}}, with the load imbalance tolerance set to 1.05 to encourage macro-counties of similar sizes.

\end{document}